\documentstyle[preprint,aps,amssymb]{revtex}
\begin{document}
\draft
\newfont{\form}{cmss10}




\def\Journal#1#2#3#4{{#1} {\bf #2}, #3 (#4)}

\def\NCA{\em Nuovo Cimento}
\def\NIM{\em Nucl. Instrum. Methods}
\def\NIMA{{\em Nucl. Instrum. Methods} A}
\def\NPB{{\em Nucl. Phys.} B}
\def\PLB{{\em Phys. Lett.}  B}
\def\PRL{\em Phys. Rev. Lett.}
\def\PRD{{\em Phys. Rev.} D}
\def\ZPC{{\em Z. Phys.} C}

\def\st{\scriptstyle}
\def\sst{\scriptscriptstyle}
\def\mco{\multicolumn}
\def\epp{\epsilon^{\prime}}
\def\vep{\varepsilon}
\def\ra{\rightarrow}
\def\ppg{\pi^+\pi^-\gamma}
\def\vp{{\bf p}}
\def\ko{K^0}
\def\kb{\bar{K^0}}
\def\al{\alpha}
\def\ab{\bar{\alpha}}
\def\be{\begin{equation}}
\def\ee{\end{equation}}
\def\bea{\begin{eqnarray}}
\def\eea{\end{eqnarray}}
\def\CPbar{\hbox{{\rm CP}\hskip-1.80em{/}}}
\def\bom#1{\mbox{\boldmath$#1$}}
\def\kap{\bom{k}}


\title{Light-front vacuum and instantons in two dimensions}

\author{A. Bassetto and F. Vian }

\address{Dipartimento di Fisica ``G. Galilei'' and INFN, Sezione di Padova,\\
via Marzolo 8, 35131 Padua, Italy}

\author{L. Griguolo}

\address{Dipartimento di Fisica ``M. Melloni'' and INFN, Gruppo Collegato di Parma,\\
viale delle Scienze, 43100 Parma, Italy}

\maketitle

\begin{abstract}
We review several aspects of Yang-Mills theory (YMT)
in two dimensions, related to its perturbative and topological properties.
Consistency between light-front and equal-time formulations is thoroughly
discussed.
\end{abstract}

\vskip 2.0truecm
\noindent
DFPD 00/TH 20

\noindent
UPRF-2000-06

\vfill\eject

\narrowtext

\section{INTRODUCTION}

Non-abelian quantum gauge theories are
still far from being satisfactorily understood.
Though some non-perturbative
features are thought to be transparent, a consistent framework in the
continuum is lacking.

Therefore one often resorts to the simplified context of
two-dimensional theories where exact solutions can sometimes be
available. In two dimensions the theory looks seemingly trivial when quantized
in the light-cone gauge (LCG) $A_-\equiv \frac{A_0-A_1}{\sqrt 2}=0.$  
As a matter of fact, in the absence of dynamical
fermions, no physical local degrees of freedom appear in the Lagrangian.

Still topological degrees of freedom occur if the theory is put
on a (partially or totally) compact manifold, whereas the simpler 
behavior on the plane enforced by the LCG condition 
entails a severe worsening in its infrared structure.
These features are related aspects of the same basic issue: even in two
dimensions ($D=2$) the theory contains some non-trivial dynamics. 
We can say that, in LCG, dynamics gets hidden
in the very singular nature of correlators at large distances
(IR singularities). 

In order to fully appreciate this point and the controversial aspects
related to it, let us briefly review the 't Hooft's
model for $QCD_2$ at large $N$, $N$ being the number of colours
\cite{thooft}.
In LCG no self-interaction occurs for the gauge fields; in the large-$N$
limit planar diagrams dominate, without quark loops. The $q\bar q$ interaction is
mediated by the exchange
\begin{equation}
\label{exch}
{\cal D}(x)=-\frac{i}{2}|x^-|\,\delta(x^+),
\end{equation}
which looks instantaneous if $x^+$ is considered as a {\it time} variable.
Eq.(\ref{exch}) is the Fourier transform of the
quantity
\begin{equation}
\label{four}
\tilde{\cal D}(k)=\frac{1}{k_-^2},
\end{equation}
the singularity at $k_-=0$ being interpreted as a Cauchy principal value.
Such an expression in turn can be derived by quantizing the theory on
the light front (at equal $x^+$), $A_+$ behaving as a constraint
\cite{noi1}.

The full set of ladder diagrams can easily be summed, leading to a 
beautiful pattern of $q\bar q$-bound states with squared masses
lying on rising Regge trajectories. This was the first evidence, 
to our knowledge, of a {\it stringy} nature of $QCD$ in its confining 
regime, reconciling dual models with a partonic field theory.

\smallskip

Still, if the theory within the same gauge choice is canonically
quantized {\it at equal times}, a different expression is obtained for
the exchange in eq.(\ref{exch})
\begin{equation}
\label{caus}
{\cal D}_{c}(x)=\frac{1}{2\pi}\,\frac{x^-}{-x^++i\epsilon x^-},
\end{equation}
and its Fourier transform
\begin{equation}
\label{fcaus}
\tilde {\cal D}_{c}(k)=\frac{1}{(k_-+i\epsilon k_+)^2},
\end{equation}
can now be interpreted as a {\it causal} Feynman propagator \cite{noi1}.

This expression, first proposed by Wu \cite{wu}, is nothing but
the restriction at $D=2$ of the prescription for the LCG vector
propagator in four dimensions suggested by Mandelstam
and Leibbrandt \cite{leibb} (ML), and derived in ref.\cite{noi}
by equal-time canonical quantization of the theory.

In dimensions higher than two, where ``physical'' degrees of freedom
are switched on (transverse ``gluons''), this causal prescription is
mandatory in order to
get correct analyticity properties, which in turn are the basis of
any consistent renormalization program \cite{noi2}.

\smallskip

When eq.(\ref{fcaus}) is used in summing the very same set of planar 
diagrams considered by 't Hooft, no rising Regge trajectories
are found in the spectrum of the $q\bar q$-system. The bound-state
integral equation looks difficult to be solved; early approximate
treatments \cite{webb} as well as a more detailed recent study 
\cite{shuva}
indicate the presence of a massless solution, with a fairly obscure 
interpretation, at least in this context. Confinement seems lost.

Then, how can it be that the causal way to treat the infrared (IR) 
singularities,
which is mandatory in higher dimensions, leads to a disastrous result
when adopted at $D=2$ ? In order to get an answer we address ourselves 
to the $q\bar q$-potential.

\section{THE WILSON LOOP}

A very convenient gauge invariant way of looking at the $q\bar q$-potential 
is to consider a rectangular Wilson loop, centered at the origin, with
sides parallel to a spatial direction and to the time direction, of length
$2L$ and $2T$ respectively
\begin{equation}
\label{wilson}
{\cal W}= {1\over N} \langle 0| {\rm Tr}\left[ {\cal T}{\cal P} 
{\rm exp} \left( ig \oint_\gamma dx^\mu A_\mu (x) \right)\right]
|0\rangle, 
\end{equation}
the symbols ${\cal T}$ and ${\cal P}$ denoting temporal ordering of
operators and colour ordering.

It is well known that the Wilson loop we have hitherto introduced
can be thought to describe the interaction of a couple of static
$q\bar q$ at the distance $2L$ from each other. 
We can turn to the Euclidean formulation replacing $T$ with $iT$. If we 
denote by ${\cal E}_0(L)$ the ground state energy of the system, we get
for large $T$
\begin{equation}
\label{ground}
{\cal W}=\exp[(4m-2{\cal E}_{0})T]
\int_{{\cal E}_{0}}^{\infty}
d{\cal E}\,\rho(L,{\cal E})\,\exp[-2T({\cal E}-{\cal E}_0)].
\end{equation} 
Unitarity requires the spectral density
$\rho(L,{\cal E})$ to be a non-negative measure. 
Then ${\cal W}$ is positive and the coefficient
of the exponential factor $\exp[(4m-2{\cal E}_{0})T]$ is a non-increasing
function of $T$.

\smallskip

We can define the 
$q\bar q$-potential as $${\cal V}(L)={\cal E}_0(L)-2m.$$
If the theory confines, ${\cal V}(L)$ is an increasing function of
the distance $L$; if at large distances the increase is linear in $L$,
namely ${\cal V}(L)\simeq 2 \sigma L,$ we obtain an area-law behaviour 
for the
leading exponent with a string tension $\sigma$. 

For $D>2$ perturbation theory is unreliable in computing
the true spectrum of the $q\bar q$-system. However, when combined
with unitarity, it puts an
intriguing constraint on the $q\bar q$-potential. To realize this point,
let us consider the formal expansion
\begin{equation}
\label{expand}
{\cal V}(L)=g^2\, {\cal V}_1(L)+g^4\, {\cal V}_2(L)+\cdots,
\end{equation}
$g$ being the QCD coupling constant.

When inserted in the expression $\exp[-2{\cal V}(L)T]$, it gives 
\begin{equation}
\label{loopex}
\exp[-2T{\cal V}]= 1-2T\bigl[g^2\, {\cal V}_1+g^4\, 
{\cal V}_2+\cdots\bigr]
+2T^2\bigl[g^4\,{\cal V}_1^2+\cdots\bigr]+\cdots.
\end{equation}
At ${\cal O}(g^4)$, the coefficient
of the leading term at large $T$ should be half the square of the term
at ${\cal O}(g^2)$. This constraint has often been used as a check
of (perturbative) gauge invariance.

Therefore, if we denote by $C_{F(A)}$ the quadratic Casimir expression for the
fundamental (adjoint) representation
of $SU(N)$ and remember that ${\cal V}_1$ is proportional to $C_{F}$,
at ${\cal O}(g^4)$ the term with the coefficient $C_{F}C_{A}$
should be subleading in the large-$T$ limit with respect to the
Abelian-like term, which is proportional to $C_{F}^2$.

Such a calculation at ${\cal O}(g^4)$ for the loop ${\cal W}$
has been performed using Feynman gauge in \cite{belluo}, with  
the number of space-time dimensions
larger than two ($D>2$). The result depends on the area ${\cal A}=4L\,T$ 
and on the dimensionless
ratio $\beta=\frac{L}{T}$. The ${\cal O}(g^2)$-term is obviously
proportional to $C_{F}$; at ${\cal O}(g^4)$ we find 
that the non-Abelian term is indeed
subleading
\begin{equation}
\label{sublead}
T^2\,{\cal V}^{na}\propto C_{F}C_{A}{\cal A}^2\, T^{4-2D}.
\end{equation}
Therefore agreement with exponentiation holds and the validity
of previous perturbative tests of gauge invariance in higher
dimensions is vindicated. 

The limit of our result when $D\to 2$ is {\it finite} and depends only
on ${\cal A}$, as expected on the basis of the invariance of the theory
in two dimensions under area-preserving diffeomorphisms. However the
non-Abelian term is no longer subleading in the limit $T\to \infty$,
as it is clear from eq.(\ref{sublead}); we get instead \cite{belluo}
\begin{equation}
\label{lead}
2T^2\,{\cal V}^{na}= C_{F}C_{A}\frac{{\cal A}^2}{16\pi^2}(1+\frac{\pi^2}{3}).
\end{equation}

We conclude that the limits $T\to \infty$ and $D\to 2$ {\it do not commute}.

This result is confirmed by a calculation of ${\cal W}$ performed
in LCG with the ML prescription for the vector propagator \cite{colf}. 
At odds with Feynman gauge where the vector
propagator is not a tempered distribution at $D=2$, in LCG the calculation
can also be performed directly in two space-time dimensions. The result
one obtains does {\it not} coincide with eq.(\ref{lead}). One gets instead
\begin{equation}
\label{lead2}
2T^2\,{\cal V}^{na}= C_{F}C_{A}\frac{{\cal A}^2}{48}.
\end{equation}

The extra term in eq.(\ref{lead}) originates from the self-energy correction
to the vector propagator. In spite of the fact that the triple vector
vertex vanishes in two dimensions in LCG, the self-energy correction 
does not. We
stress that this ``anomaly-like'' contribution is not a pathology 
of LCG, it is needed
to comply with  Feynman 
gauge.

Perturbation theory is {\it discontinuous} at $D=2$.
We conclude that the perturbative result, no matter what
gauge one adopts, conflicts with unitarity in two dimensions.

\smallskip

Taking advantage of the invariance under area-preserving diffeomorphisms
in dimensions $D=2$, Staudacher and Krauth \cite{sk} were
able to generalize our ${\cal O}(g^4)$ result (eq.(\ref{lead2})) by fully resumming 
the perturbative series.
In the Euclidean formulation, which is allowed as the causal
propagator can be Wick-rotated, and with a particular choice of the
contour (a circumference), 
they get
\begin{equation}
\label{krauth}
{\cal W}({\cal A})
=\frac{1}{N}\exp\Big[- \frac { g^2\cal A}{4}\Big]L^{(1)}_{N-1}
\Bigl(\frac{g^2 {\cal A}}{2}\Bigr),
\end{equation}
the function $L^{(1)}_{N-1}$ being a  Laguerre polynomial.

This result can be further generalized to a loop winding $n$-times
around the countour
\begin{equation}
\label{risultatino}
{\cal W}=\frac{1}{N}\exp\left[-\frac{g^2 {\cal A} \,n^2}4\right]\,
L_{N-1}^{(1)}\left(\frac{g^2 {\cal A} \,n^2}{2} \right).
\end{equation}
 
>From eq.(\ref{krauth}) one immediately realizes that, 
for even values of $N$, the result is no
longer positive in the large-$T$ limit. Moreover in the 't Hooft's
limit $N \to \infty$ with $g^2 N=2\hat g^2$ fixed, the string
tension vanishes and eq.(\ref{krauth}) becomes
\begin{equation}
\label{bessel}
{\cal W} \to \frac{1}{\sqrt{\hat g^2 {\cal A}}}
J_1(2\sqrt{\hat g^2 {\cal A}}),
\end{equation}
$J_1$ being the usual Bessel function.
Confinement is lost.

This explains the failure of the Wu's approach in getting a bound state
spectrum lying on rising Regge trajectories in the large-$N$ limit.

\smallskip

However in LCG 
the theory can also be quantized on the {\em light-front} (at equal $x^{+}$);
with such a choice, in pure YMT and just in two dimensions,
no dynamical degrees of freedom occur as the non
vanishing component of the vector field does not propagate,
but rather gives rise to an instantaneous (in $x^{+}$) Coulomb-like
interaction (see eq.(\ref{exch})). 

Only planar diagrams contribute to the Wilson loop ${\cal W}$ for
any value of $N$, thanks to the ``instantaneous'' nature of such
an exchange; the perturbative series can be easily resummed,
leading to the result (for imaginary time)
\begin{equation}
\label{area}
{\cal W}({\cal A})
=\exp\Big[- \frac { g^2N\cal A}{4}\Big],
\end{equation}
to be compared with eq.(\ref{krauth}).

Not only is this result in complete agreement with the exponentiation 
required by unitarity; it also exhibits, in the 't Hooft's limit
$N \to \infty$ with $g^2 N=2\hat g^2$ fixed, confinement with a finite string
tension $\sigma=\frac{\hat g^2}{2}$.
This explains the success of 't Hooft's approach in computing the spectrum
of the $q\bar q$ bound states.
The deep reason of this good behaviour lies in the absence of ghosts
in this formulation; however there is no smooth way of deriving it
from any acceptable gauge choice in
higher dimension $(D>2)$. Moreover the confinement exhibited at this stage
is, in a sense, trivial, being shared by $QED_2$.

We end up with two basically different results for the {\it same} model
and with the {\it same} gauge choice (LCG), according to the different ways
in which IR singularities are regularized.
Moreover we are confronted with the following paradox:
the prescription which is mandatory in dimensions $D>2$ is the
one which fails at $D=2$. What is the meaning (if any) of eq.(\ref{krauth})?

\section{THE GEOMETRICAL APPROACH}

In order to understand this point, it is worthwhile to study the problem
on a compact two-dimensional manifold; possible IR
singularities will be automatically regularized in a gauge invariant way.
For simplicity, we choose
the sphere $S^2$. We also consider the slightly simpler
case of the group $U(N)$. 
On $S^2$ we envisage a smooth non self-intersecting closed contour 
and a loop winding around it a number $n$ of times.
We call $A$ the total area of the sphere,
which eventually will be sent to $\infty$, whereas ${\cal A}$ will be
the area ``inside'' the loop we keep finite in this limit.

Our starting point is the well-known heat-kernel 
expressions \cite{migdal} of a non self-intersecting Wilson loop
for a pure $U(N)$ YMT on a sphere with area $A$
\begin{eqnarray}
\label{wilsonz}
&&{\cal W}_{n}(A,{\cal A})={1\over {\cal Z}(A)N}
\sum_{R,S} d_{R}d_{S}
\exp\left[-{{g^2 {\cal A}}\over 2}C_2(R)-{{g^2 (A-{\cal A})}\over 2}
C_2(S)\right]\nonumber \\
&&\times \int dU {\rm Tr}[U^{n}]\chi_{R}(U) \chi_{S}^{\dagger}(U),
\end{eqnarray}
$d_{R\,(S)}$ being the dimension of the irreducible representation $R(S)$ of
$U(N)$; $C_2(R)$ ($C_2(S)$) is the quadratic Casimir expression, 
the integral in (\ref{wilsonz}) is over the
$U(N)$ group manifold while $\chi_{R(S)}$ is the character of the group
element $U$ in the $R\,(S)$ representation. ${\cal Z}(A)$ is the partition 
function of the theory, its explicit form being easily obtained from 
${\cal W}_{0}(A,{\cal A})=1$.

We write eq.(\ref{wilsonz}) explicitly for $N>1$ and $n>0$ in the form
\begin{eqnarray}
\label{wilsonp}
&&{\cal W}_{n}(A,{\cal A})=\frac{1}{{\cal Z}(A)}
\sum_{m_i=-\infty}^{+\infty}\Delta(m_1,...,m_N)
\Delta(m_1+n,m_2,...,m_N)\nonumber \\
&&\times \exp\left[-\frac{g^2A}{4}\sum_{i=1}^N (m_i)^2
 \right]
\exp\left [-\frac{g^2 n}{4}(A-{\cal A})(n+2m_1)\right].
\end{eqnarray}
We have described the generic irreducible representation by means
of the set of integers $m_{i}=(m_1,...,m_{N})$, related to the
Young tableaux, in terms of which
we get
\begin{equation}
\label{casimiri}
C_2(R)=\frac{N}{24}(N^2-1)+\frac{1}{2}\sum_{i=1}^{N}(m_{i}-\frac{N-1}{2})^2,
\qquad d_{R}=\Delta(m_1,...,m_{N}).
\end{equation}
$\Delta$ is the Vandermonde determinant and
the integration in eq.(\ref{wilsonz})
has been performed explicitly, using the well-known formula for the 
characters in terms of the set $m_{i}$ and taking symmetry into account.

>From eq.(\ref{wilsonp}) it is possible to derive, for $n=1$ and 
in the large-$A$ 
decompactification limit, precisely the expression 
(\ref{area}) we obtained by resumming the perturbative series
in the 't Hooft's approach. 
This is a remarkable result as it has now been
derived in a purely geometrical way without even fixing a gauge.
Actually, in the decompactification limit $A\to \infty$ at fixed
${\cal A}$, from eq.(\ref{wilsonp}) one gets the following expression
for any value of $n$ and $N$ \cite{bgv}
\begin{eqnarray}
\label{wilmorenice}  
&&{\cal W}_{n}({\cal A};N)=\frac{1}{n N} \, \exp\Bigl[-\frac{g^2 {\cal A}}{4}
\, n(N+n-1)\Bigr]\nonumber \\
&&\times\!\!\sum_{k=0}\!\frac{(-1)^k\,\Gamma(N+n-k)}{k!\,\Gamma(N-k)\Gamma(n-k)}
\exp\Bigl[\frac{g^2 {\cal A}}{2}\,n\,k \Bigr].
\end{eqnarray}
 
We notice 
that when $n>1$ the simple abelian-like exponentiation is lost. In other 
words the theory starts feeling its non-abelian nature as the appearance 
of different ``string tensions'' makes clear.
The winding number $n$ probes its colour content.
The related light-front vacuum, although simpler than the one in the
equal-time quantization, cannot be considered trivial any longer.

Eq.(\ref{wilmorenice}) exhibits an interesting symmetry under the exchange of
$N$ and $n$. 
More precisely, we have that
\begin{equation}
\label{dual}
{\cal W}_{n}({\cal A};N)={\cal W}_{N}(\tilde{\cal A};n),\qquad\qquad
\tilde{\cal A} = \frac {n}{N} \,{\cal A} \,,
\end{equation}
a relation that is far from being trivial, involving an unexpected
interplay between the geometrical and the algebraic structure of the
theory \cite{bgv}.

Looking at eq.(\ref{dual}), the abelian-like exponentiation for $U(N)$
when $n=1$ appears to be related to the $U(1)$ loop with $N$
windings, the ``genuine'' triviality of Maxwell theory providing the
expected behaviour for the string tension. Moreover we notice the
intriguing feature that the large-$N$ limit (with $n$ fixed) is
equivalent to the limit in which an infinite number of windings is
considered with vanishing rescaled loop area. Alternatively, this 
rescaling could be thought to affect the
coupling constant $g^2 \to \frac{n}{N} g^2$.

>From eq.(\ref{wilmorenice}), in the limit $N\to \infty$, one can recover
the Kazakov-Kostov result \cite{kaza}
\begin{equation}
\label{vecchia}
{\cal W}_{n}({\cal A};\infty) = \frac1n 
L^{(1)}_{n-1}(\frac{\hat{g}^2 {\cal A}n}{2}) \,
\exp \Bigl[ -\frac{\hat{g}^2 {\cal A} n}4\Bigr].
\end{equation}

Now, using eq.(\ref{dual}) we are able to perfom another limit, namely $n\to
\infty$ with fixed $n^2\,{\cal A}$  
\begin{equation}
\label{granden}
\lim_{n\to\infty} {\cal W}_{n}({\cal A};N)  =
\frac 1N \,
L^{(1)}_{N-1}\Bigl( g^2 {\cal A}\,n^2 /2 \Bigr)
 \exp \Bigl[ -\frac{g^2 {\cal A}\, n^2}4\Bigr]  \,.
\end{equation}

We remark that this large-$n$ result reproduces the resummation of the
perturbative series (for any $n$) (eq.(\ref{risultatino})) in the {\it causal}
formulation of the theory. 

\smallskip

We go back to the exact expression we have found on the sphere 
for the Wilson loop (eq.\ref{wilsonp}).
As first noted by Witten \cite{Witte}, it is possible to
represent ${\cal W}_{n}(A,{\cal A})$ (and consequently ${\cal Z}(A)$) 
as a sum over
instable instantons, where each instanton contribution is 
associated to a finite,
but not trivial, perturbative expansion. The easiest way to see it, is 
to perform a Poisson resummation
\begin{eqnarray}
\label{poisson}
&&\sum_{m_{i}=-\infty}^{+\infty}\!F(m_1,...,m_{N})=
\sum_{f_{i}=-\infty}^{+\infty}\!\tilde{F}(f_1,...,f_{N}),\\
\tilde{F}(f_1,...,f_{N})&=&\int_{-\infty}^{+\infty}dz_1...dz_{N}
F(z_1,...,z_{N})
\exp \left[2\pi i(z_1 f_1+...+z_{N}f_{N})\right]\nonumber
\end{eqnarray}
in eq.(\ref{wilsonp}).
One gets
\begin{eqnarray}
\label{istanti}
&&{\cal W}_{n}(A,{\cal A})=\frac{1}{{\cal Z}(A)}\exp\left[\frac{g^2n^2
(A-2{\cal A})^2}{16 A}\right]\times\nonumber \\
&&\sum_{f_{i}=-\infty}^{+\infty}
\exp\left[-S_{inst}(f_{i})\right]W(f_1,...,f_{N})
\exp\left[-2 \pi in f_{1}\frac{A-{\cal A}}{A}\right],
\end{eqnarray}
where
\begin{equation}
\label{quantita`}
S_{inst}(f_{i})=\frac{4\pi^2}{g^2 A}\sum_{i=1}^{N}f_{i}^2,
\end{equation}
and
\begin{eqnarray}
\label{zetawu}
&&W(f_1,...,f_{N})=\int
_{-\infty}^{+\infty}dz_1...dz_{N}
\exp\left[-\frac{1}{g^2A}
\sum_{i=1}^{N}z_{i}^2\right]\exp\left(\frac{inz_{1}}
{2}\right)\times \nonumber\\
&&\Delta(z_1-2\pi\tilde f_1,...,z_N-2\pi f_N)\,\,
\Delta(z_1+2\pi\tilde f_1,...,z_N+2\pi f_N),
\end{eqnarray}
with $$\tilde f_1=f_1+\frac{ig^2n}{8\pi}(A-2{\cal A}).$$

These formulae have a nice interpretation in terms of instantons.
Indeed, on $S^2$, there are non trivial solutions of the Yang-Mills equation,
labelled by the set of integers $f_{i}=(f_1,...,f_{N})$ 
\begin{equation}
\label{monopolo}
{\cal A}_{\mu}(x)={\rm Diag}\left(f_1{\cal A}_{\mu}^{0}(x), f_2 {\cal A}_{\mu}^{0}(x), 
\ldots, f_N{\cal A}_{\mu}^{0}(x)\right)
\end{equation}
where ${\cal A}_{\mu}^{0}(x)={\cal A}_{\mu}^{0}(\theta, \phi)$ is the Dirac
monopole potential,
$${\cal A}_{\theta}^{0}(\theta, \phi)=0 , \quad\   {\cal A}_{\phi}^{0}
(\theta, \phi)={1-\cos \theta\over 2}, $$
$\theta$ and $\phi $ being spherical coordinates on $S^{2}$. 
The term $\exp\left[-2 \pi in f_{1}\frac{A-{\cal A}}{A}\right]$ 
in eq.(\ref{istanti}) corresponds to the classical contribution of such
field configurations to the Wilson loop.

Only the zero instanton contribution should be obtainable 
by means of a genuine perturbative calculation. Therefore  
in the following we single out
the zero-instanton contribution ($f_{q}=0$, $\forall q$) to the Wilson loop 
in eq.(\ref{istanti}), obviously normalized to the zero instanton 
partition function \cite{bg}.

The equation,
after a suitable rescaling, becomes
\begin{equation}
\label{zeroinst}
{\cal W}_{n}^{(0)}(A,{\cal A})=\frac{1}{{\cal Z}^{(0)}(A)}
\exp\left[\frac{g^2n^2
(A-2{\cal A})^2}{16 A}\right]
W_{1}(0,...,0)
\end{equation}
with
\begin{eqnarray}
\label{wzeroinst}
&&W_{1}(0,...,0)=\int
_{-\infty}^{+\infty}dz_1...dz_{N}
\exp\left[-\frac{1}{2}
\sum_{i=1}^{N}z_{i}^2\right]\exp\left(\frac{in\sqrt {g^2A}z_{1}}
{2\sqrt 2}\right)\\
&&\times \Delta(z_1-\frac{in}{4}\sqrt {\frac{2g^2}{A}}
(A-2{\cal A}),\cdots,z_N)
\Delta(z_1+\frac{in}{4}\sqrt{\frac{2g^2}{A}}(A-2{\cal A}),\cdots,z_N)
\nonumber.
\end{eqnarray}

The two Vandermonde determinants can be expressed in terms of
Hermite polynomials \cite{gross} and then expanded in the usual way.
The integrations over $z_2,...,z_{N}$ can be performed, taking
the orthogonality of the polynomials into account; we get 
\begin{eqnarray}
\label{integrata}
&&{\cal W}_{n}^{(0)}(A,{\cal A})=
\exp\left[\frac{g^2n^2
(A-2{\cal A})^2}{16 A}\right]
\prod_{n=0}^{N}\frac{1}{n!}\prod_{k=2}^{N}(j_{k}-1)!
\frac{\varepsilon^{j_1...j_{N}}\varepsilon_{j_1...j_{N}}}{{\cal Z}^{(0)}(A)}
\nonumber \\
&&\int_{-\infty}^{+\infty}dz_1
\exp\left[-\frac{1}{2}
z_{1}^2\right]
\exp\left(\frac{in\sqrt {g^2A}z_{1}}
{2\sqrt 2}\right)
He_{j_1-1}(z_{1+})He_{j_1-1}(z_{1-}),
\end{eqnarray}
where
\begin{equation}
\label{zetapm}
z_{1\pm}=z_1\pm\frac{in}{4}\sqrt{\frac{2g^2}{A}}(A-2{\cal A}).
\end{equation}
Integration over $z_1$ finally gives
\begin{equation}
\label{risultato}
{\cal W}_{n}^{(0)}(A,{\cal A})=
\frac 1N \,
L^{(1)}_{N-1}\Bigl(\frac{g^2 {\cal A}(A-{\cal A})\,n^2}{2A} \Bigr)
\exp \Bigl[ -\frac{g^2 {\cal A}(A-{\cal A})\, n^2}{4A}\Bigr]  \,.
\end{equation}

At this point we remark that, in the decompactification limit $A\to
\infty$, ${\cal A}$ fixed, the quantity in the equation above {\it exactly}
coincides, for any value of $N$, with eq.(\ref{risultatino}), which
was derived following completely different considerations.
We recall indeed that eq.(\ref{risultatino}) was obtained 
by a full resummation of the perturbative expansion of the Wilson loop
in terms of causal Yang-Mills propagators in LCG.
Its meaning is elucidated by
noting that it just represents the zero-instanton contribution to
the Wilson loop, a genuinely perturbative quantity \cite{bg}.

In turn it also coincides with the expression of the exact result
in the large-$n$ limit, keeping fixed the value of $n^2 {\cal A}$
(eq.(\ref{granden})). This feature can be understood if we remember
that instantons have a finite size; therefore small loops are 
essentially blind to them \cite{bgv}.

If the perturbative result has been correctly interpreted, it should be related 
to the local behaviour of the theory: in particular it should be possible
to derive it starting 
from any topology, when the decompactification limit is eventually performed. 
In \cite{gri} the zero-instanton contribution to a homologically trivial 
Wilson loop on the torus $T^2$ has been computed: the Poisson resummation is harder 
there and a larger number of classical solutions complicates the geometrical 
structure. In spite of the different complexity, the zero-instanton contribution, 
in the decompactification limit, still coincides with the perturbative result, 
as expected.

\section{THE \kap -SECTORS}

It was first noticed by Witten \cite{wittheta} 
that two-dimensional Yang-Mills
theory and two-dimensional QCD with adjoint matter do possess
$k$-sectors. 
We consider $SU(N)$ as the gauge group: since Yang-Mills fields
transform in the adjoint representation, the true local symmetry is
the quotient of $SU(N)$ by its center, $Z_N$. A standard result in
homotopy theory tells us that the quotient is not simply
connected, the first homotopy group being $\Pi_1(SU(N)/Z_N)=Z_N.$
This result is of particular relevance for the vacuum structure of a
two-dimensional gauge theory: in the case at hand we have
exactly  $N$ inequivalent quantizations, parametrized by 
a single integer $k$, taking the values $k=0,1,..,N-1$. 
Concerning the pure $SU(N)$ Yang-Mills theory, the  explicit solution 
when $k$-states are taken into account was presented in
Ref.~\cite{canad}:  their 
main result, the heat-kernel propagator on the cylinder, allows to compute 
partition functions and Wilson loops on any two-dimensional compact 
surface, therefore generalizing  the well-known Migdal's 
solution \cite{migdal} to $k$-sectors. Wilson loops, in this case,  
strongly depend on $k$: for a non self-intersecting loop we have, on the plane,
\begin{eqnarray}
&&{\cal W}_{k} (A)=\frac1{N^2-1} \left[ 1+
\frac{k N (N+2)(N-k)}{(k+1)(N-k+1)}
e^{-\frac{g^2 A}2 \,(N+1)}
\right. \\
&+& \left.
\frac{(N+1)(N-k-1)}{k+1}
e^{-\frac{g^2 A}2 \,(N-k)}+ 
\frac{(N+1)(k-1)}{N-k+1}
e^{-\frac{g^2 A}2 \,k}\right].
\end{eqnarray}
This result can be obtained starting from the true $SU(N)/Z_N$ theory on the 
sphere \cite{bgv2}, in the decompactification limit, or 
directly on the plane, using the procedure of \cite{paniak}, working with 
$SU(N)$ and simulating the $k$-sectors with a Wilson loop at infinity in the 
$k-$fundamental representation. The very same result can be obtained 
through a perturbative resummation \cite{bgv2} with 't Hooft potential and 
the $k$-loop at infinity. On the other hand we expect that the truly perturbative 
physics ignore the existence of the $k$ parameter: by Poisson-resumming the
result on 
the sphere for $SU(N)/Z_N$, we arrive to an instanton representation 
different from the $SU(N)$ case. Contribution from $N$ 
inequivalent classes of instantons ensues, 
with instanton numbers generalized to rational values by the effect of $k$. 
The zero-instanton limit does not depend on $k$ and still reproduces the WML 
computation (in the decompactification limit) {\it without} the loop at infinity 
\cite{bgv2}
\begin{eqnarray}
\label{wilzero}
&&{\cal W}_{k}^{(0)}(A)=\frac{1}{N+1}+\frac{N}{{\cal Z}
\,(N+1)}\int_{-\infty}^{+\infty}dz_1\ldots dz_N
\exp \left[ -\frac{1}{2}\sum_{j=1}^{N}z_j^2 \right] \times\nonumber \\
&&\exp\Bigl[ig(z_1-z_2)
\sqrt\frac{A}{2}\Bigr] \Delta^2(z_1,\ldots,z_N),
\end{eqnarray}
where ${\cal Z}=\int {\cal D}F\,
\exp(-\frac{1}{2}{\rm Tr}F^2).$
In the presence of a $k$-loop at infinity, the WML computation, although 
coinciding with the zero-instanton limit of the quantum average of two nested 
($k$ and adjoint) loops, does depend on $k$. 
We conclude that ${\it only}$ for the complete theory on the 
plane (i.e. full-instanton resummed and then decompactified) the equivalence 
between $k$-sectors and theories with $k$-fundamental Wilson loops at
infinity  holds.

\vfill\eject

\end{document}